\documentclass[conference]{IEEEtran}
\IEEEoverridecommandlockouts
\usepackage{amsmath,amssymb,amsfonts}
\usepackage{algorithm}
\usepackage{algpseudocode}
\usepackage{graphicx}
\usepackage{textcomp}
\usepackage{xcolor}
\usepackage{hyperref}
\usepackage{acro}
\usepackage{siunitx}
\usepackage{fancyhdr}
\usepackage{booktabs}
\usepackage[style=ieee,minnames=1,maxbibnames=6]{biblatex}
\DeclareAcronym{hpc}{
  short = HPC ,
  long  = high-performance computing
}
\DeclareAcronym{hep}{
  short = HEP ,
  long  = high-energy physics
}
\DeclareAcronym{dag}{
  short = DAG ,
  long  = directed acyclic graph
}
\DeclareAcronym{sbnd}{
  short = SBND,
  long  = Short-Baseline Near Detector
}
\DeclareAcronym{bnb}{
  short = BNB,
  long  = Booster Neutrino Beam
}
\DeclareAcronym{lartpc}{
  short = LArTPC,
  long  = liquid argon time projection chamber
}
\DeclareAcronym{alcf}{
  short = ALCF,
  long  = Argonne Leadership Computing Facility
}
\DeclareAcronym{des}{
  short = DES,
  long= discrete-event simulation
}
\DeclareAcronym{fifo}{
  short = FIFO,
  long  = first-in-first-out
}

\newcommand{\parsl}{\textsc{Parsl}}
\newcommand{\sbnparsl}{\textsc{sbn\_parsl}}
\newcommand{\larsoft}{\textsc{LArSoft}}
\newcommand{\wirecell}{\textsc{WireCell}}
\newcommand{\geant}{\textsc{Geant4}}
\newcommand{\genie}{\textsc{Genie}}
\newcommand{\pandora}{\textsc{Pandora}}

\newcommand{\figref}[1]{Fig.~\ref{fig:#1}}

\newcommand{\eqnref}[1]{Eq.~\ref{eq:#1}}
\newcommand{\tabref}[1]{Tab.~\ref{tab:#1}}
\newcommand{\algoref}[1]{Alg.~\ref{alg:#1}}

\def\BibTeX{{\rm B\kern-.05em{\sc i\kern-.025em b}\kern-.08em
    T\kern-.1667em\lower.7ex\hbox{E}\kern-.125emX}}
\begin{document}

\title{Real-Time Optimization of Simulation and Data Processing Pipelines for Experiments on Exascale Computing Platforms}

\author{\IEEEauthorblockN{Thomas Wester}
\IEEEauthorblockA{\textit{Enrico Fermi Institute} \\
\textit{University of Chicago}\\
Chicago, IL, USA \\
0000-0001-6668-7595}
\and
\IEEEauthorblockN{Harikrishna Tummalapalli}
\IEEEauthorblockA{\textit{Argonne Leadership Computing Facility} \\
\textit{Argonne National Laboratory}\\
Lemont, IL, USA \\
0000-0002-1055-3130}
\and
\IEEEauthorblockN{Christine M. Simpson}
\IEEEauthorblockA{\textit{Argonne Leadership Computing Facility} \\
\textit{Argonne National Laboratory}\\
Lemont, IL, USA \\
0000-0001-9985-1814}
}

\maketitle

\thispagestyle{fancy}
\lhead{}
\rhead{}
\chead{}
\lfoot{\footnotesize{
SC26 Workshops, November 15-20, 2026, Chicago, Illinois, USA
\newline 979-8-3195-1221-5/26/\$31.00 \copyright 2026 IEEE}}
\rfoot{}
\cfoot{}
\renewcommand{\headrulewidth}{0pt}
\renewcommand{\footrulewidth}{0pt}

\begin{abstract}
Current and next-generation large-scale experiments, especially in high-energy physics and cosmology, involve increasing volumes of data that require exascale computing power to analyze and model. Exascale machines present challenges for workflow scaling and efficient resource usage. We propose a multi-stage task scheduling approach for optimizing resource usage for these workflows. We demonstrate task scheduling for an example Monte Carlo simulation pipeline from the Short-Baseline Near Detector (SBND), and present theoretical and numerical tools for modeling task scheduling in these workflows that can be used to identify optimal scheduling strategies and choose resources for a given problem size.
\end{abstract}


\section{\label{sec:intro}Introduction}

Current and next-generation scientific experiments will collect petabytes of raw instrument data~\cite{dunetdr}. Data-intensive analysis patterns have been identified as a key pillar of the workloads that national computational facilities will need to support to realize the scientific impact of many upcoming experiments~\cite{IRI_blueprint}, and exascale \ac{hpc} systems provide the necessary computational resources to process this volume of data on the timescales necessary for typical data analysis cycles. However, extreme-scale systems are often many times bigger than resources experimentalists are accustomed to using near their experiments and therefore pose a challenge in porting and optimizing pipelines.

Experimental data are typically organized into discrete units, e.g., events or objects, which must be processed through staged computational pipelines in order to extract high-level quantities of interest~\cite{LSSTSimSkySurvey}. Optimal resource utilization of these computational pipelines is often linked to finding the optimal schedule. Pipeline scheduling has been extensively studied~\cite{benoit2013survey}. However, these studies often consider deterministic workloads, where the task durations are predictable. In practice, advance knowledge is not always available: real experimental data and Monte Carlo simulations often contain statistical fluctuations that affect the run times of each processing stage~\cite{LSSTparsl}. For example, a particle detector may observe a random number of particles in a given event, each of which must be processed through a pattern recognition algorithm. Static scheduling algorithms often perform poorly on these stochastic workloads~\cite{zheng2012monte}. The problem is exacerbated by the heterogeneity of modern \ac{hpc} systems.

In this work, we show how a real-time task scheduling strategy can improve resource utilization and reduce the makespan, or wall time, for stochastic pipeline workflows on production \ac{hpc} systems. Additionally, we provide theoretical and numerical tools to estimate makespan, resource utilization, and other performance metrics for general workflow \acp{dag} with random task durations. Our scheduling strategy divides computational pipelines into two stages: a short prediction stage which is used to estimate the run time of a second, longer stage. This strategy is broadly applicable to science domains which process variable amounts of data in multiple steps, and we show its direct application to a \ac{hep} Monte Carlo simulation workflow.

The paper is structured as follows: we first describe relevant theoretical expressions used for optimally scheduling tasks with unknown run times on \ac{hpc} systems. We then present a workflow simulator which can be used to accurately estimate makespan for different workflow configurations. Finally, we compare the theoretical and simulated results to a real \ac{hep} application for simulating neutrino interactions running on an \ac{hpc} system. The consistency of the three approaches validates that a two-stage pipeline with a dynamic scheduling policy reduces the makespan and improves worker occupancy on \ac{hpc} systems and provides a tool for exploring scaling and optimization of workloads as described in the last section.

\section{\label{sec:theory}Scheduling Theory}

Optimal pipeline scheduling is well studied for particular cases in the literature~\cite{benoit2009scheduling, benoit2013survey}, and it is well known that minimizing the makespan is NP-hard for an arbitrary configuration of task run times~\cite{benoit2009scheduling}. Consequently, scheduling strategies are employed to assign computational tasks from pipelines to \ac{hpc} resources in a particular order, e.g., based on task characteristics or at random.

We consider submitting a workflow consisting of $M$ tasks to an \ac{hpc} job or ``block'' for processing. An \ac{hpc} block consists of $N$ workers which will be in use until an unknown makespan, or wall time, $T$, at which point the final task finishes. The tasks are assigned to workers from a shared, perfect queue, i.e., when one worker finishes, it begins processing the next available task immediately unless there are no more tasks in the queue, at which point it is idle. The performance of the job can be measured by the makespan, $T$, and the idle time, $I$. Shorter $T$ and smaller $I$ represent better usage of the block.

Workflows can be optimized trivially to minimize $T$ by setting $N=M$. However, when the variance of task run times is non-zero, this configuration introduces a maximum inefficiency as measured by $I$, since many workers are idle while longer tasks finish. On the other extreme, setting $N=1$ minimizes $I$ but is impractical as the makespan $T$ is maximized instead. \ac{hpc} users must choose an acceptable $T$ and $I$ given the number of tasks and the run-time variance of their particular real-world application.

To derive the relationship of $T$ and $I$ for different scheduling strategies, we assume tasks have random run times from a distribution $q(t)$ with mean $\mu$ and variance $\sigma^2$. We consider two scheduling scenarios: either the task run times are not known in advance of the block and are randomly assigned to workers, or the task run times are known in advance and can be ordered with a scheduling strategy. Since task run-time distributions may have long tails, it is useful to discuss these quantities as a function of the fraction of completed tasks, $p$, i.e., the $T$ or $I$ at which a fraction $p$ of the work is done, which we denote as $T_{p}$ and $I_{p}$.

\subsection{Unknown Run Times (Unordered)}

This scheduling algorithm assumes no prior knowledge about the tasks, such that the tasks are taken from the queue at random. The block can be divided into two phases: a bulk phase when there are tasks in the queue, and a draining phase which starts at a time $t_{\mathrm{drain}}$ when the final task is removed from the queue, which is approximately $(M-N)\mu/N$, i.e., the time at which $M-N$ tasks have completed. To find $T_p$, we consider the number of tasks completed during each phase. In the bulk phase, before the queue is empty, all workers are busy. In the draining phase, which begins when the final task is assigned to a worker at $t_{\mathrm{drain}}$, we denote the remaining run time of each task by $\tau$. The distribution of remaining run times, $r(\tau)$, can be obtained by assuming the remaining run times are uniformly distributed with respect to $t_{\mathrm{drain}}$\footnote{This assumption should be considered optimistic and holds only when $M\gg N$ (in the limiting case where $M=N$, $t_{\mathrm{drain}}$ starts immediately so the run times are not uniformly distributed, and $r(\tau)$ would be exactly the run time distribution, $q(t)$, rather than a residual time distribution).}. This assumption also has the consequence that the probability of a task being present on a worker is proportional to its run time, $t$, i.e., longer-duration tasks are more likely to be running at $t_{\mathrm{drain}}$. Under this assumption, the remaining tasks have run times from a distribution given by

\begin{equation}\label{eq:qdrain}
    q_{\mathrm{drain}}(t) = \frac{t q(t)}{\mu},
\end{equation}

\noindent and the conditional probability of $\tau$ for a given task with run time $t$ is $1/t$. $r(\tau)$ then considers all tasks with at least $\tau$ seconds remaining via the integral over all $t$,

\begin{equation}
    r(\tau) = \int_{\tau}^{\infty} \frac{q_{\mathrm{drain}}(t) }{t} dt.
\end{equation}

\noindent Substituting \eqnref{qdrain}, we find

\begin{equation}
    r(\tau) = \frac{1}{\mu} \int_{\tau}^{\infty}  q(t)dt = \frac{1-Q(\tau)}{\mu},
\end{equation}

\noindent where $Q(t)$ is the cumulative distribution function of $q(t)$. The number of tasks completed within a time $\tau$ after $t_{\mathrm{drain}}$ is then given by integrating $r$,

\begin{equation}
m_{\mathrm{drain}} = N \int_0^{\tau} r(\delta) d\delta \equiv NR(\tau),
\end{equation}

\noindent where $R(\tau)$ is the cumulative distribution of remaining run times. The makespan $T_p$ can be expressed as a function of the completed fraction of tasks as

\begin{equation}\label{eq:tprandom}
    T_p = \begin{cases}
    \frac{pM \mu}{N} & m_{\mathrm{bulk}} / M \geq p,\\
    \frac{m_{\mathrm{bulk}} \mu}{N}+ R^{-1}\left(\frac{m_{\mathrm{drain}}}{N}\right), & m_{\mathrm{bulk}} / M < p,\\
    \end{cases}
\end{equation}

\noindent where $R^{-1}(x)$ is the inverse of $R$, $m_{\mathrm{bulk}} \equiv M-N$, and $m_{\mathrm{drain}} = p M - m_{\mathrm{bulk}}$ is the number of tasks that must complete during the draining phase to reach $p$. The idle time, $I_p$, can then be expressed as

\begin{equation}
    I_p = N T_p - pM\mu.
\end{equation}

\subsection{Known Run Times (Ordered)}

When run times are known, a standard heuristic-based scheduling policy is to prioritize either longest or shortest tasks. Since the residual duration of the block is proportional to the task durations at the end of the queue, scheduling the longest tasks first is the better strategy. In this case, we also have a bulk phase and draining phase. To find $T$ and $I$, we need only consider the final task to be submitted to a worker. By construction, the final task is assigned to the first available worker after all other tasks are assigned, so it runs on the worker with the minimum sum of task run times. In the worst-case scenario, this last task starts at the moment all other tasks finish and extends $T$ by its run time, giving an upper bound on the completion of the final task. When $M\gg N$, this sets the makespan\footnote{If the longest-running task is still running at the time the last task is assigned to a worker, its run time could extend the makespan instead. However, this case is atypical when $M\gg N$.}

\begin{equation}
T_{\mathrm{LF}} \leq \frac{1}{N}\sum_{i=1}^{M-1} t_i + t_{M}.
\end{equation}

\noindent Here, the $t_i$ are the run times of the other tasks, and $t_M$ is the final (shortest run time) task. From this, we also obtain the idle time, $I$, as

\begin{equation}
    I_{\mathrm{LF}} \leq  N t_M,
\end{equation}

\noindent which only scales with the number of workers allocated during the block. Importantly, for known, randomly distributed run times scheduled according to a longest-first policy, $t_{M}$ decreases as $M$ increases, since the final task is the shortest by definition. For a random distribution of task run times, this means $I$ decreases for large $M$, which is the opposite trend of the unknown run-time case.

We obtain $T_p$ and $I_p$ for this case by defining the number of tasks $k$ required to reach the desired completion fraction $p$ once the queue is empty. Since the tasks are run in order, $T$ is extended by the run time of the $k^\mathrm{th}$ task, which is, in general, found by evaluating $Q^{-1}(k/M)$. The idle time $I_p$ is found by integrating the run times over the bulk and draining phases,

\begin{equation}
    W_{\mathrm{bulk}} = M \int^1_{N/M} Q^{-1} (t) dt,
\end{equation}

\noindent i.e., the total run time of the longest $M-N$ tasks, assuming the draining phase starts with $N$ tasks remaining. Similarly,

\begin{equation}\label{eq:wdrain}
    W_{\mathrm{drain}} = M \int^{k/M}_0 Q^{-1} (t) dt.
\end{equation}

\noindent We then find $I_p = N T_p - W_{\mathrm{bulk}} - W_{\mathrm{drain}}$.

\bigskip

The expressions derived in this section rely on idealized assumptions and are therefore optimistic estimates for the different scheduling algorithms. As we have seen, information about the task run times allows for a more favorable reduction in both the makespan and idle time, resulting in better utilization of \ac{hpc} systems.

\section{\label{sec:simulator}Scheduling Simulator}

\begin{figure}
    \centering
    \includegraphics[width=0.75\linewidth]{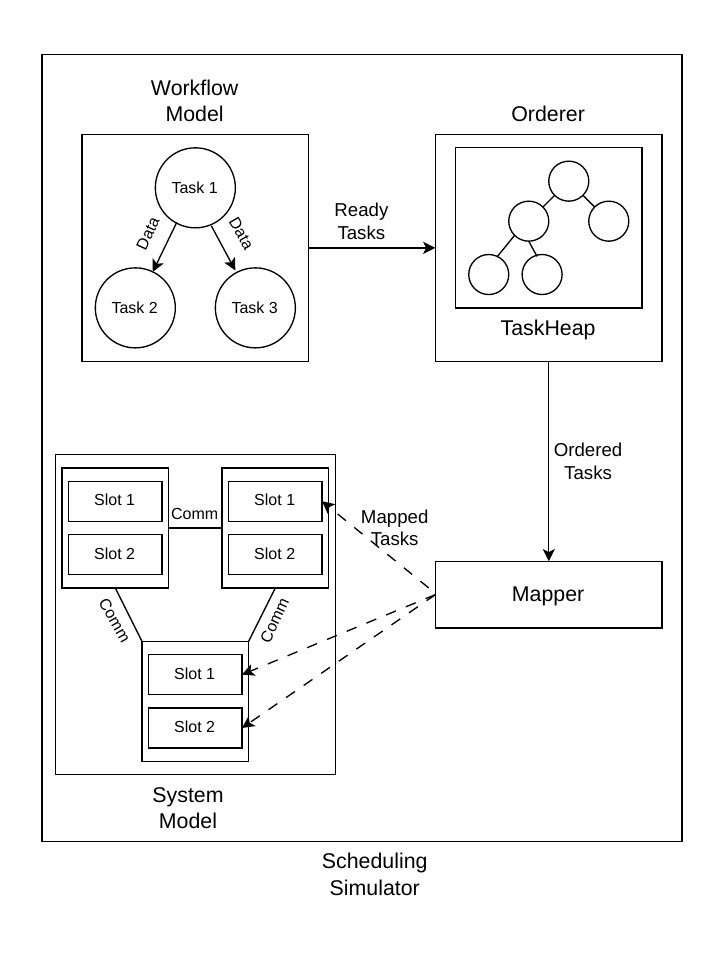}
    \vspace{-15pt}
    \caption{Illustration of the workflow scheduling simulator.}
    \label{fig:illustrate_simulator}
\end{figure}

Since the theoretical estimates of the makespan $T$ and worker idle time $I$ require simplifying assumptions, we developed a scheduling simulator to numerically estimate these quantities. The simulator extends the pipeline scheduling framework described in \cite{benoit2013survey} to general \acp{dag} and decomposes scheduling into a series of ordering and mapping steps. An architectural overview is shown in \figref{illustrate_simulator}.

The simulator consists of four main components: \texttt{WorkflowModel}, \texttt{TaskOrderer}, \texttt{Mapper}, and \texttt{SystemModel}. The \texttt{WorkflowModel} holds a set of \texttt{Workflow} objects, each representing a workflow as a \ac{dag}. Each task in the workflow (\texttt{WorkflowTask}) is characterized by a \texttt{compute\_cost} and a \texttt{data\_size} parameter, representing the computational cost and the output data size of the task, respectively. Once all dependencies of a task are resolved, it is marked as ready for execution. Ready tasks are inserted into a priority queue (\texttt{TaskHeap}) using a key defined by the chosen ordering policy. We provide several built-in policies, including \ac{fifo}, shortest-first, and longest-first. The ordered tasks are then mapped onto the compute system. The \texttt{SystemModel} consists of compute nodes (\texttt{ComputeNode}), each containing a collection of compute slots (\texttt{ComputeSlot}). Each slot is characterized by a compute power parameter. The inter-node network topology is represented as a \texttt{networkx}~\cite{hagberg2007exploring} graph, where each edge carries a bandwidth parameter.

The simulator drives these components through a \ac{des} loop, presented in \algoref{des}. The loop maintains the current simulation time step, $t$, an event priority queue, $\mathcal{Q}$, and a waiting priority queue, $\mathcal{W}$. At initialization, all ready tasks are obtained from the workflow model, $\mathcal{M}$, and inserted into $\mathcal{W}$ with priorities assigned by the orderer, $\mathcal{O}$. These tasks are then scheduled onto the system, $\mathcal{S}$, via the mapper, $\mathcal{F}$. Within the \textsc{Schedule} procedure, the completion time $t_c$ of each allocated task is computed and inserted into the event queue along with the task, $\tau$, and its allocated slots, $\mathbf{s}$. At each iteration, the earliest completion event is extracted from $\mathcal{Q}$ and the simulation clock is advanced to $t_c$. The corresponding resources are released in $\mathcal{S}$, and the task is marked as completed in $\mathcal{M}$. Any newly ready successors are inserted into $\mathcal{W}$ and scheduled. The loop terminates when $\mathcal{Q}$ is empty.

\begin{algorithm}[t]
\small
\caption{Discrete-event workflow scheduling simulation}
\label{alg:des}
\begin{algorithmic}[1]
\Require Workflow model $\mathcal{M}$, system model $\mathcal{S}$, mapper $\mathcal{F}$, orderer $\mathcal{O}$
\Ensure Makespan $T$
\Statex
\State $t \gets 0$; \enspace $\mathcal{Q} \gets \emptyset$; \enspace $\mathcal{W} \gets \emptyset$
\Statex
\For{each $\tau \in \mathcal{M}.\textsc{ReadyTasks}()$}
    \State Insert $\tau$ into $\mathcal{W}$ with priority $\mathcal{O}(\tau)$
\EndFor
\State \Call{Schedule}{$\mathcal{W},\mathcal{S},\mathcal{F},\mathcal{Q},t$}
\Statex
\While{$\mathcal{Q} \neq \emptyset$}
    \State $(t, \tau, \mathbf{s}) \gets $ \Call{ExtractMin}{$\mathcal{Q}$}
    \State $\mathcal{S}.\textsc{Release}(\mathbf{s})$
    \State $\mathcal{M}.\textsc{MarkCompleted}(\tau, t)$
    \For{each $v \in \mathcal{M}.\textsc{ReadyTasks}(\tau)$}
        \State Insert $v$ into $\mathcal{W}$ with priority $\mathcal{O}(v)$
    \EndFor
    \State \Call{Schedule}{$\mathcal{W},\mathcal{S},\mathcal{F},\mathcal{Q},t$}
\EndWhile
\State \Return $t$
\Statex
\Procedure{Schedule}{$\mathcal{W},\mathcal{S},\mathcal{F},\mathcal{Q},t$}
    \If{$\mathcal{W} = \emptyset$ \textbf{or} $\mathcal{S}.\textsc{free} = 0$} \Return \EndIf
    \For{each $(\tau, \mathbf{s}) \in \mathcal{F}(\mathcal{W}, \mathcal{S})$}
        \State $t_c \gets t + c(\tau) \,/\, p(\mathbf{s})$
        \State \Call{Insert}{$\mathcal{Q}, (t_c, \tau, \mathbf{s})$}
    \EndFor
\EndProcedure
\end{algorithmic}
\end{algorithm}

In the following sections we validate the simulator by comparing to a production workflow on an \ac{hpc} system, and use it to investigate the performance of our scheduling strategy at different node counts and task duration variance.

\section{\label{sec:hep}High-Energy Physics Workflow}

\begin{figure}
    \centering
\includegraphics[width=1.0\linewidth]{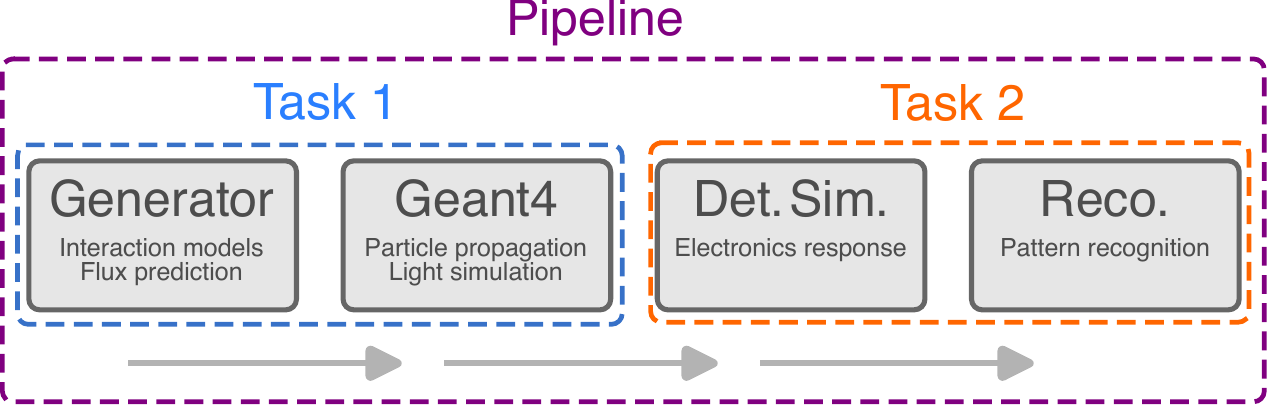}    \caption{The \acs{hep} Monte Carlo processing pipeline discussed in the text. The pipeline has four stages that each depend on the previous stage. The scheduling strategies considered either execute all the stages from a pipeline in sequence on a worker, or separately schedule stages grouped into tasks 1 and 2, potentially on different workers.}
    \label{fig:hep_workflow}
\end{figure}

\ac{hep} experiments typically consist of a detector which produces a data stream of triggered events. Events are a record of the state of the detector during readout windows, where the activity in the detector largely depends on the particle multiplicities and energies present during the window. Events undergo several pre-processing steps before being ultimately processed through pattern recognition algorithms to identify high-level features of interest such as particle tracks, momentum estimates, etc. Most \ac{hep} experiments also compare data to Monte Carlo simulations in which particle interactions are generated according to a particular physics model, and events are simulated according to a realistic description of the detector geometry.

A typical \ac{hep} Monte Carlo processing pipeline involves a number of stages: generation of initial particle multiplicities and kinematics, propagation of the generated particles through a detector medium, simulated detector readout of the propagated particles, and ``reconstruction,'' i.e., running signal processing and pattern recognition algorithms on the detector readout. Simulation pipelines can be described as a \ac{dag}; the stages of each pipeline are run in sequence, and the results of each stage provide the inputs for the next. Typical simulation sizes are in the millions of events for modern \ac{hep} experiments~\cite{dunetdr}. To produce a large number of events quickly, the stages of each \ac{dag} are distributed across parallelized computational resources, each processing a small portion of the total work while respecting ordered dependencies.

We compare the efficacy of two scheduling algorithms on a simulation workflow from the \ac{sbnd} experiment at Fermilab~\cite{sbn2015,sbn2019}. \ac{sbnd} is a \qty{112}{\tonne} active mass \ac{lartpc} placed in Fermilab's \ac{bnb}. \ac{sbnd} has been collecting data since 2024 and has recorded millions of neutrino interactions. \ac{sbnd}'s simulation software is based on \larsoft~\cite{larsoft2013}, a configurable, module-based library collection with specific utilities for \acp{lartpc}. A typical \larsoft{} workflow consists of many copies of a pipeline with several stages. The first stage runs the \genie{} event generator~\cite{genie2009} to simulate neutrino interactions according to specified physics models, followed by \geant~\cite{geant2002} to simulate the propagation of the particles through liquid argon. The next stage simulates the detector readout using a realistic simulation of the electric field configuration in the detector and an electronics response model implemented with \wirecell~\cite{wirecell2022}. Finally, the last stage produces high-level information for analysis by reconstructing the simulated events with the \pandora~\cite{pandora2015} pattern recognition algorithm. The steps of the workflow are illustrated in \figref{hep_workflow}.

\begin{figure}
    \centering
\includegraphics[width=0.9\linewidth]{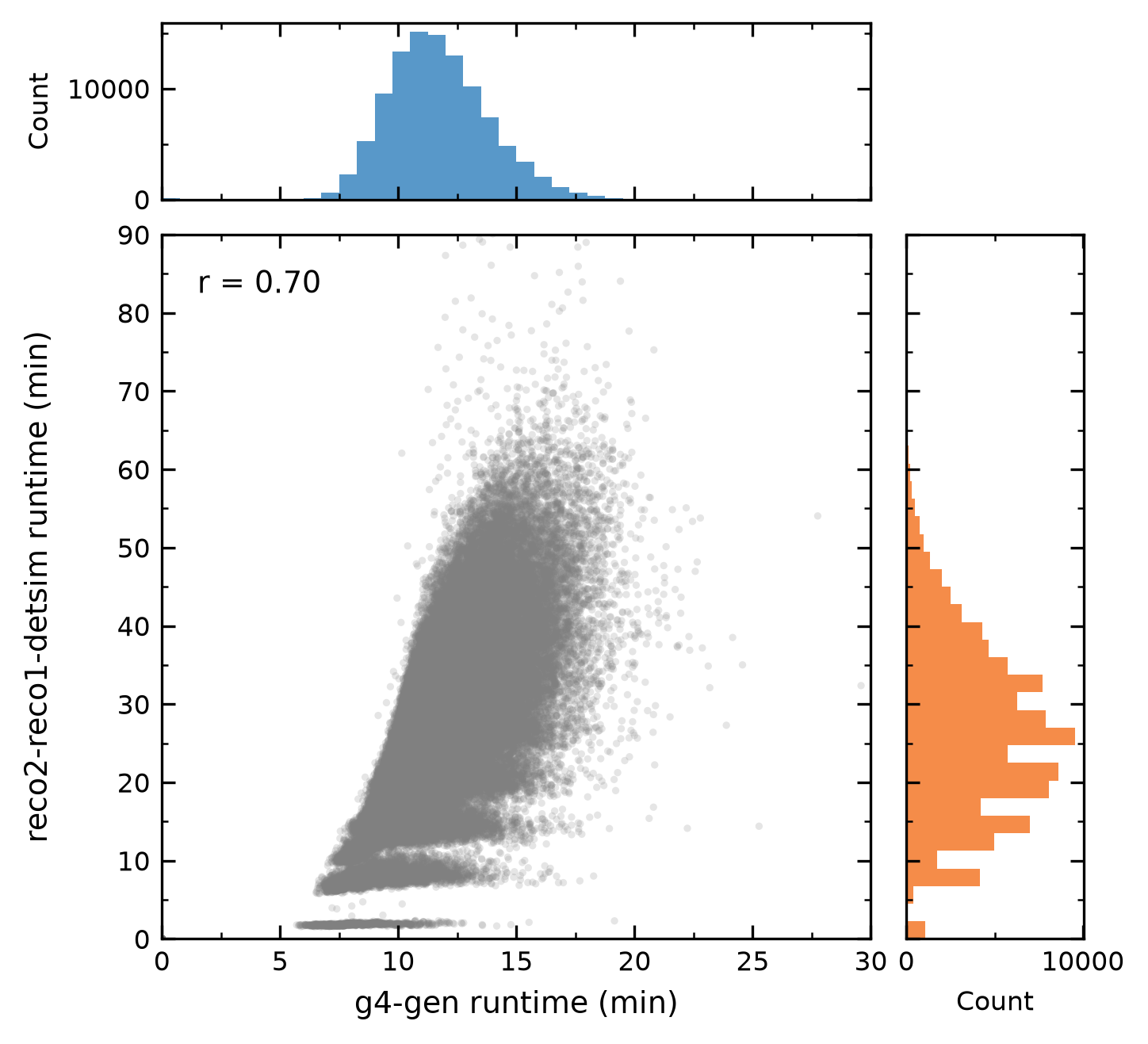}    
\vspace{-10pt}
\caption{Task time distribution and correlation between the two stages of the \acs{hep} workflow. The banding pattern is a feature of the discrete number of events generated in the first stage.}
    \label{fig:correlation}
\end{figure}

Because each pipeline begins with a generation step that produces a random number of particles, downstream steps are expected to vary in run time. However, there is a strong correlation between the run times of the early stages of the pipeline and later stages. The correlation is illustrated in \figref{correlation} for the \larsoft{} workflow described above, shown for \num{100000} independent pipelines. Using this correlation in a multi-stage scheduling strategy allows us to modify the ordering of later stages as earlier stages complete to prioritize stages with longer estimated times first in a dynamic queue.

\begin{figure*}
    \centering
\includegraphics[width=0.75\linewidth]{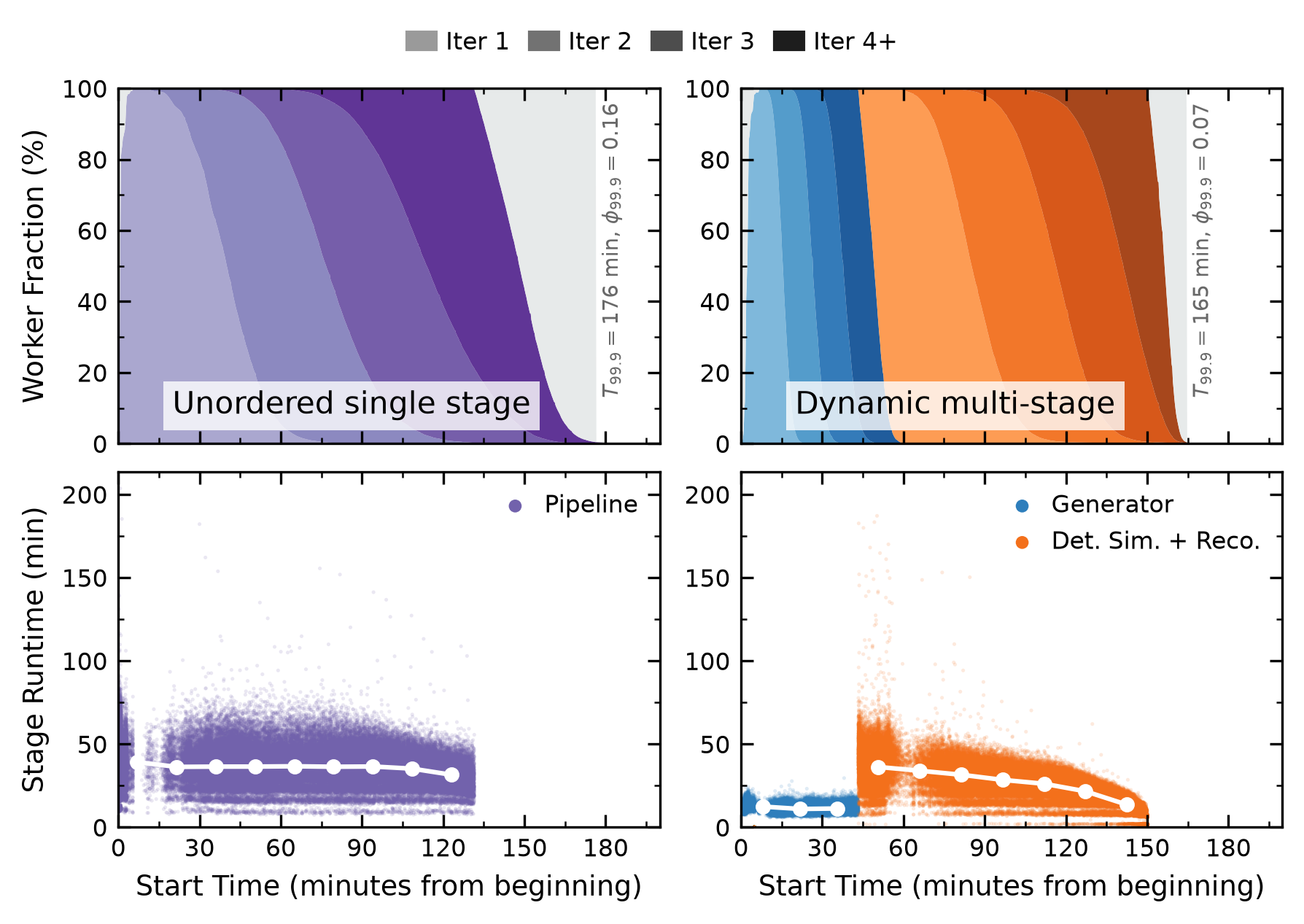}    
\vspace{-10pt}
\caption{Worker occupancy versus block start time (top panels) and run time versus start time for tasks (bottom panels) for two scheduling strategies on Aurora. The worker count is \num{26112} and the pipeline count is \num{104448}. The left column shows the \textit{unordered single-stage} case, while the right column shows the \textit{dynamic multi-stage} case. Colors correspond to task types, either an entire pipeline (unordered single-stage) or generator and downstream stages (dynamic multi-stage). Color shade indicates the iteration number, i.e., how many tasks of each type have already run on each worker. In the lower panels, the white lines show the median stage run times. The discontinuity around 10 minutes in the unordered single-stage case is due to task start time synchronization at the start of the job, and the banding pattern results from the discrete, randomized number of events simulated in each pipeline.}
    \label{fig:aurora_results}
\end{figure*}

Scheduling tests of the case study application were conducted on Aurora at the \ac{alcf}. Aurora is a \num{10624}-node HPE Cray EX system. Each node has two Intel Xeon CPU Max 9470C processors, of which \num{102} of \num{104} cores are available for science, with \qty{1}{\tera\byte} of DDR5 memory and \qty{128}{\giga\byte} of HBM, along with six Intel Data Center GPU Max Series GPUs. We developed the package \sbnparsl, a wrapper around the \parsl{} library~\cite{parsl2019}, to bundle and execute large numbers of \acp{dag} representing \ac{lartpc} simulation and data processing pipelines on \ac{hpc} systems. Internally, \sbnparsl{} leverages \parsl's \texttt{HighThroughputExecutor} to distribute worker processes across nodes in multi-node block allocations. We ran \num{102} workers per node, one per available core. GPUs were not used in this test for simplicity, but we note that the scheduling strategies presented here could be generalized for other queue-based workflows that include GPUs (e.g.,~\cite{icecube2024}). To support different task scheduling priorities within \parsl, \sbnparsl{} modifies \parsl's \texttt{DataFlowKernel}, which prioritizes tasks.

The following task scheduling strategies are tested:

\begin{enumerate}
    \item \textit{Unordered single-stage} --- Schedules each pipeline \ac{dag} on a single worker, which executes the stages sequentially, effectively reducing the \ac{dag} to a single task.
    \item \textit{Dynamic multi-stage} --- \acp{dag} are split into two execution units as shown in \figref{hep_workflow}. The generator stages are prioritized ahead of the downstream stages. As the generator stages complete, their corresponding simulation/reconstruction stages are given an updated priority according to the run length of the generator stage.
\end{enumerate}

Workflow timelines are shown in \figref{aurora_results}. Top panels show the worker occupancy, as measured by task start and end times recorded by \sbnparsl. Bottom panels show the stage run times as a function of their start time. In the unordered single-stage case, only one computational task processes the entire pipeline. In the dynamic multi-stage case, the generator stages run first and set the priorities of the downstream simulation stages; we see the median stage run time decrease at later start times, as is characteristic of longest-first scheduling.

\section{\label{sec:results}Results}

We tested the scheduling approaches for the \ac{hep} workflow at \num{256}-node scale ($N=\num{26112}$ workers), with a generous wall time request to fully capture the long run-time tail produced by a small number of pipelines. The total number of pipelines was $M=\num{104448}$, so that $M/N=4$. We measured the makespan to complete different fractions of the pipelines, $T_{p}$, and the idle fraction at that time, $\phi_{p} = I_p / (N \times T_p)$. We refer to its complement, $1-\phi_p$, as the worker occupancy.

\begin{figure*}
    \centering
\includegraphics[width=0.75\linewidth]{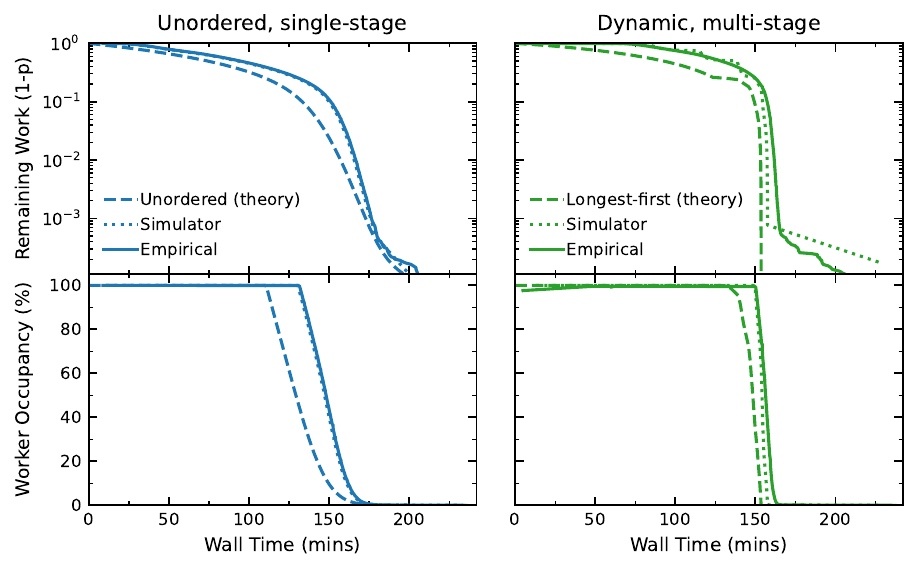}
\vspace{-10pt}
\caption{Comparison of scheduling results from this paper. Upper panels show the remaining fraction of work to complete ($1-p$) and lower panels show the worker occupancy. The left and right columns show the unordered single-stage and dynamic multi-stage scheduling strategies, respectively.}
    \label{fig:three_compare}
\end{figure*}

We compared the \ac{hep} workflow results to the theoretical results and to the simulator in \figref{three_compare} and tabulated comparison metrics in \tabref{results}. The theoretical results were calculated from Eqs.~\ref{eq:tprandom}--\ref{eq:wdrain}, using an empirical task run-time distribution, $q(t)$, from the timings obtained from the empirical case, and using numerical methods to obtain $Q(t)$ and its inverse from this distribution. We set $M=\num{104448}$ and $N=\num{26112}$ for consistency with the empirical case. For the simulator, we used the task run times from the production workflow to build the corresponding workflow model. For the unordered single-stage strategy, we used the \ac{fifo} orderer, and for the dynamic multi-stage strategy, we implemented a custom orderer that uses the tuple of stage number and negative of task duration to set the task priority.

\begin{table}
    \centering
    \caption{\normalfont\upshape Expected makespans, $T_p$, and idle fractions, $\phi_p$, for the theoretical, simulator, and empirical \acs{hep} workflow cases. The theory values are idealized estimates, calculated using the task run time distribution from the empirical case and with assumptions discussed in the text; they are expected to underestimate $T_p$.}
    \label{tab:results}
    \begin{tabular*}{\linewidth}{@{\extracolsep{\fill}}
      S[table-format=2.2]
      *{3}{
        S[table-format=3.1, round-mode=places, round-precision=1]
        S[table-format=2.1, round-mode=places, round-precision=1]
      }
    @{}}
    \toprule
    & \multicolumn{2}{c}{Theory (Limit)} & \multicolumn{2}{c}{Simulator} & \multicolumn{2}{c}{Empirical} \\
    \cmidrule{2-3} \cmidrule{4-5} \cmidrule{6-7}
    {$p$} & {$T_p$} & {$\phi_p$} & {$T_p$} & {$\phi_p$} & {$T_p$} & {$\phi_p$} \\
    {(\%)} & {(\unit{\minute})} & {(\%)} & {(\unit{\minute})} & {(\%)} & {(\unit{\minute})} & {(\%)} \\
    \midrule
    \multicolumn{7}{c}{\textbf{Unordered single-stage}} \\
    \midrule
    99 & 157.20 &  7.08 & 166.18 & 11.33 & 166.70 & 11.04 \\
    99.9 & 172.77 & 14.69 & 179.62 & 17.87 & 176.55 & 15.92 \\
    99.99 & 199.05 & 25.89 & 234.72 & 37.13 & 212.12 & 30.00 \\
    \midrule
    \multicolumn{7}{c}{\textbf{Dynamic multi-stage}} \\
    \midrule
      99 & 153.32 &  4.52 & 157.40 &  2.19 & 161.83 &  4.10 \\
    99.9 & 153.62 &  4.03 & 157.63 &  2.33 & 164.58 &  6.68 \\
    99.99 & 153.64 &  3.98 & 228.58 & 32.61 & 214.67 & 27.66 \\
    \bottomrule
    \end{tabular*}
    \end{table}

In \figref{three_compare}, we see that the dynamic multi-stage strategy has noticeably better performance in terms of smaller $T_{p}$ and smaller idle fraction $\phi_{p}$ than the unordered case, confirming an improvement in performance by using the predicted run time distribution for ordering the second pipeline stages. The results are confirmed numerically in \tabref{results}, with the disclaimer that $\phi_{99.99}$ and $T_{99.99}$ are measured with approximately $M\times (1-0.9999) \approx 10$ tasks only, and are therefore subject to large statistical variance.

We find the real-world workflow follows the expected trend from the theoretical results. In \figref{three_compare}, the empirical and simulator worker occupancies have similar slopes to the theoretical predictions. The lag in makespan, as indicated by larger remaining work fractions at equivalent makespans for the empirical and simulator cases in the upper panels, is a direct consequence of the simplifying assumptions made in the equations. For example, the assumption that the tasks are uniformly distributed at the moment the queue empties is optimistic, and  holds only for $M/N \gg 1$, whereas our measurements are at $M/N = 4$. We see a similar pattern in the dynamic multi-stage case compared to the longest-first theoretical limit, though both the simulator and the measurement reproduce the sharp drop in remaining work and worker occupancy characteristic of this scheduling strategy.

\section{\label{sec:explore}Pipeline Simulations: Properties and Scaling}

Using the scheduling simulator which we have demonstrated provides high-fidelity reconstructions of scheduling of the \ac{sbnd} workflow, we test several variations of the workflow, varying node count and run time variance. We show how the theory and simulation tool described here can help inform optimized runs at larger scales on Aurora and other systems.

For this part of the study, we include a small adaptation to the simulator.  Rather than adopt the empirical run times from the case study runs, we use the mean and variance of the task times presented in \figref{correlation} to generate collections of estimated task times with high fidelity to the science workflow.  For the unordered single-stage case this just requires drawing task times from a single normal distribution; for the dynamic multi-stage case, we draw task times for the two stages from independent normal distributions and then couple them in pipelines by sorting them and matching tasks in sorted order. This gives results that match the empirical telemetry in \figref{three_compare}. We can then explore some scenarios of optimizing for makespan and worker occupancy.

\begin{figure}
    \centering
\includegraphics[width=0.78\linewidth]{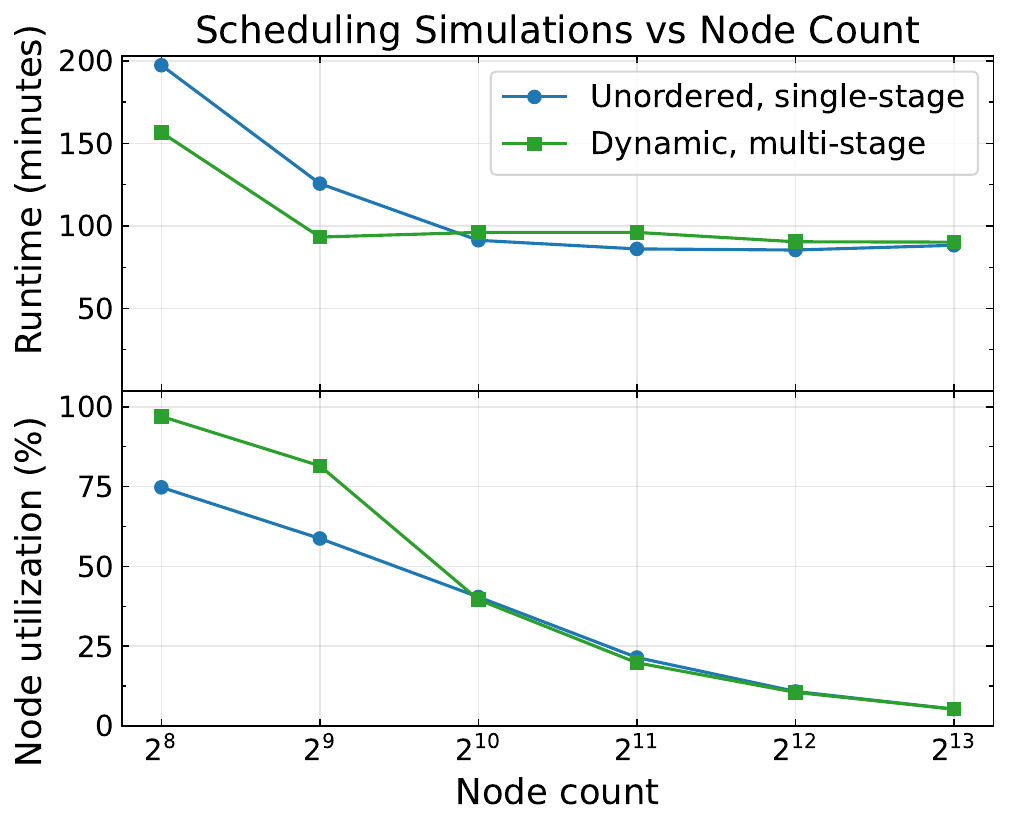}
\vspace{-10pt}
\caption{Strong scaling tests run with the scheduling simulator for task run times generated from the run-time distributions measured in the \ac{sbnd} case study.  These simulations assume \num{102} workers per node and simulate a fixed number of pipelines (\num{104448}).}
    \label{fig:scaling_sims}
    \end{figure}

First, we show simulations of strong scaling. We fix the number of pipelines simulated to be the same as the number processed in the empirical \ac{hep} workflow runs. We show the worker occupancy in \figref{scaling_sims} for the unordered single-stage and dynamic multi-stage scheduling strategies. As the node count increases, the worker occupancy drops while the makespan flattens.  This indicates an optimal node count for this pipeline and workload size where the minimum makespan is achieved with minimal loss of worker occupancy.

\begin{figure}
    \centering
\includegraphics[width=0.78\linewidth]{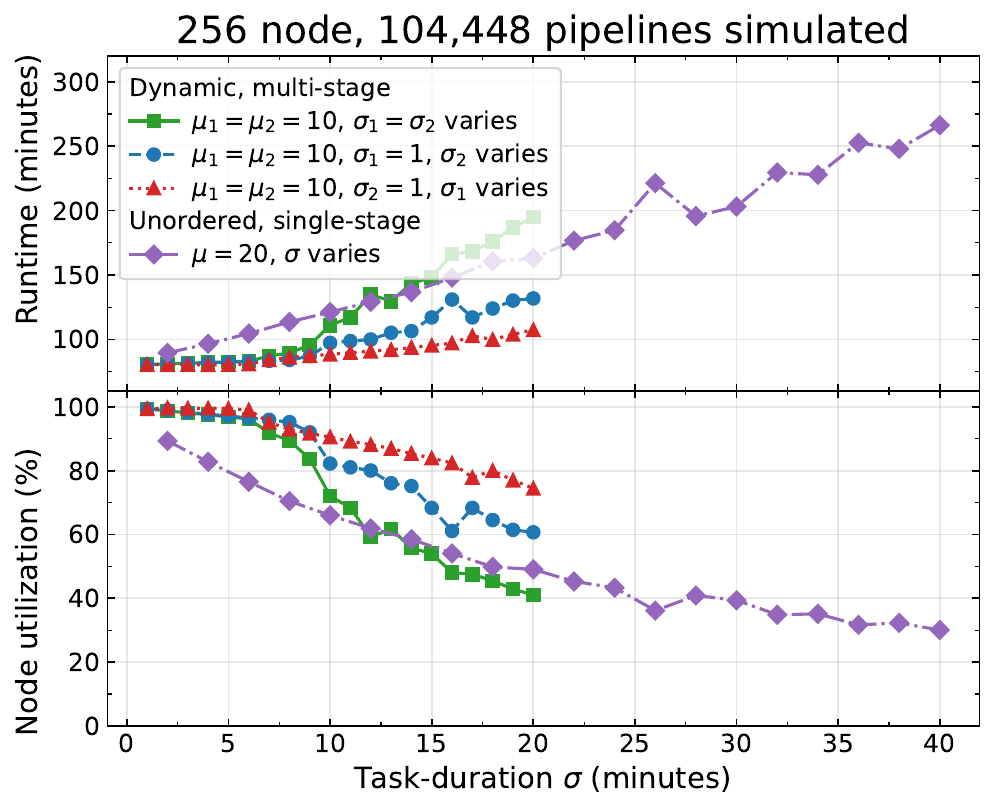}
\vspace{-10pt}
\caption{Tests exploring differing task distributions with the scheduling simulator.  Three cases of a two-stage pipeline are shown where the mean task run time for both stages is \num{10} minutes and the standard deviation in the task run times of different stages is allowed to vary.  Also shown is a version of the unordered single-stage case where the mean task run time is \num{20} minutes and the standard deviation of the task run times differs.}
\label{fig:sigma_var_sims}
\end{figure}

Second, we conduct tests where we fix the node count to \num{256} as used in the empirical \ac{hep} workflow runs, and increase the standard deviation, $\sigma$, of task run times (see \figref{sigma_var_sims}).  
For the dynamic multi-stage pipeline, we consider three variants: both stage variances increase together; stage 1 is fixed and stage 2 increases; and stage 2 is fixed and stage 1 increases.  For the unordered single-stage pipeline case, we fix the mean of the tasks to be twice the mean run time of the single stages in the dynamic multi-stage case and steadily increase $\sigma$.

The performance of the dynamic multi-stage case is far less sensitive to task run-time variance than the unordered single-stage case. Increasing the variance of both task run times together gives the worst results, and at sufficiently high variance it converges to the unordered single-stage case.  In the case where only one task stage run time increases in variance, the dynamic multi-stage case remains more performant than the unordered single-stage case. However, the case where the early stage is fixed in variance and the downstream stage varies tends to do worse.

\section{\label{sec:conclusions}Conclusions}

Optimal scheduling is necessary at exascale: our results show a high-variance distribution of task run times can lead to large idle worker times and increased makespans. As an example, in cases where approximately \num{15}\% of node compute time is left idle, as we observed at our test scale, a substantial number of node-hours can be wasted.  If the workload were increased relative to node count in such a way as to fix the makespan, at \num{2048} nodes this would be approximately \num{1000} node-hours wasted for the \ac{hep} pipeline we tested. This number can be reduced to hundreds of node-hours instead using the dynamic multi-stage scheduling strategy.

By partitioning pipelines into multiple stages, and leveraging the correlation between the stages to inform task priority, we have shown it is possible to optimize task submission. We validated this policy against a robust theoretical and simulator-based framework for assessing optimal scheduling of pipeline-based workflows on \ac{hpc} systems, and against a production Monte Carlo workflow from the \ac{sbnd} experiment.

We designed an extensible workflow simulator that can be used to develop and investigate scheduling algorithms for general workflow \acp{dag}, with both deterministic and stochastic task durations. We validated the simulator against real runs on Aurora at the \ac{alcf}, and used it to explore properties of the workflow. If properties of the pipeline vary (e.g., task run times or number of stages), the scheduling problem becomes more complex. However, the scheduling simulator we employ can help experimentalists make decisions on how to deploy pipelines if the task run-time distribution can be measured.

For this study we did not explore the efficiency of our task launcher, \parsl.  \figref{aurora_results} shows there is a small but noticeable launching inefficiency in the \num{256}-node tests we present here.  Comparisons of \parsl{} with other workflow tools have shown that task launching latency becomes an issue as node counts increase and will be an aspect of future study~\cite{ELdraft}. We also only validated and ran tests on a single resource type (CPUs).  Future work will consider mixed resource types (CPUs and GPUs). While per-task performance can improve on GPUs, the total makespan of workflows may not necessarily, in certain circumstances, be improved by the inclusion of GPUs, which are less numerous on compute nodes. 

The importance of data-intensive compute patterns for experimental science has been noted as a priority for some time \cite{IRI_blueprint}.  This study attempts to wrestle with the challenges inherent to the scale needed to deploy these patterns for a representative case from an in-production \ac{hep} experiment.  More studies of this type will help the experimental community make progress on delivering science from these experiments.

\section*{Acknowledgment}
An award of computer time was provided by the ASCR Leadership Computing Challenge (ALCC) program. This research used resources of the Argonne Leadership Computing Facility, which is a U.S. Department of Energy Office of Science User Facility operated under contract DE-AC02-06CH11357.

\printbibliography

\end{document}